\documentclass[reprint,floatfix,superscriptaddress,amsmath,amssymb,aps,prx]{revtex4-2}
\usepackage{graphicx}
\usepackage{placeins}

\begin{document}

\title{Neural Network Fractional Quantum Hall States with Learnable Orbital Pair Matrix Factors: Balancing Physical Priors and Flexibility}

\author{Yubing Qian}
\affiliation{School of Physics, Peking University}%
\affiliation{ByteDance Seed}

\author{Weiluo Ren}
\email{renweiluo@bytedance.com}
\affiliation{ByteDance Seed}

\author{Ji Chen}
\email{ji.chen@pku.edu.cn}
\affiliation{School of Physics, Peking University}
\affiliation{Interdisciplinary Institute of Light-Element Quantum Materials and Research Center for Light-Element Advanced Materials, Peking University}
\affiliation{State Key Laboratory of Artificial Microstructure and Mesoscopic Physics and Frontiers Science Center for Nano-Optoelectronics, Peking University}

\begin{abstract}
    Neural network fractional quantum Hall (FQH) states offer a promising alternative for capturing Landau level (LL) mixing and for understanding FQH physics, yet the incorporation of physical priors has so far been limited to the minimal exact constraints built into the ansatz.
    Working in spherical geometry, we show that the network's flexibility suffers both when only minimal additional physical priors are supplied and when priors such as flux attachment are imposed rigidly.
    In the former case, the monopole harmonic product orbital (MHPO) ansatz is difficult to optimize, whereas in the latter, the expressivity of the neural network is restricted.
    We therefore propose the orbital pair matrix factor MHPO (MHPO-M) ansatz, which balances physical priors against flexibility by not only replacing the fixed Laughlin structure with learnable pair factors but also making them dependent on both the electron configurations and the orbitals.
    Across $\nu=1/5$, $1/3$, and $2/5$, MHPO-M achieves the lowest variational energy among the tested variants, and a factorized orbital readout based on it enables scaling to 35 electrons, with energies competitive with trends inferred from three-LL exact-diagonalization results.
    These results demonstrate that physical priors can be appropriately incorporated without sacrificing the flexibility of neural network wavefunctions, opening a promising path toward accurate, scalable modeling of FQH systems with LL mixing.
\end{abstract}

\maketitle

\section{Introduction}

The fractional quantum Hall (FQH) effect arises from strong electron--electron interactions in partially filled Landau levels, producing topologically ordered phases with fractional excitations~\cite{stormer_fractional_1999, laughlin_anomalous_1983}.  
Coulomb interactions mix higher Landau levels (LLs)~\cite{sodemann_landau_2013}, inducing Wigner-crystal phase transitions~\cite{goldman_evidence_1990, thiebaut_fractional_2015}, spin transitions~\cite{eisenstein_evidence_1989}, and novel non-Abelian FQH phases~\cite{luhman_observation_2008, wu_braiding_2014}.  
However, traditional numerical methods for modeling FQH states face inherent limitations in addressing LL mixing.
Exact diagonalization (ED) is restricted primarily to the lowest Landau level (LLL), with limited extensions to higher levels~\cite{regnault_evidence_2017, yoshioka_effect_1984}. Density matrix renormalization group can access up to five LLs but faces challenges near critical points~\cite{feiguin_density_2008, zaletel_infinite_2015}. Fixed-phase diffusion Monte Carlo incorporates infinitely many LLs but requires a precise phase choice and lacks explicit wavefunction access~\cite{ortiz_fp-dmc_1993, zhao_crystallization_2018, zhao_composite_2023}.

Recently, deep learning methods have emerged as a fresh route for exploring FQH states, offering a direct and accurate treatment of LL mixing~\cite{qian_deephall_2025, teng_solving_2025, fadon_torus-net_2026, abouelkomsan_first-principles_2026, xue_low-variance_2026, graham_non-abelian_2026}.
These approaches represent the wavefunction with powerful neural networks and optimize it variationally through quantum Monte Carlo, pioneered by Carleo and Troyer~\cite{carleo_solving_2017}.
Such neural network FQH (NNFQH) states primarily employ neural network ansatzes adapted from studies of molecular systems~\cite{pfau_ferminet_2020, glehn_psiformer_2023}.
In molecular studies, once the exact fermionic structure is enforced, learning the remaining wavefunction structure variationally is preferable to imposing detailed priors from approximate chemistry~\cite{pfau_ferminet_2020, glehn_psiformer_2023, gerard_gold-standard_2022}.
Whether this conclusion extends to the NNFQH context, however, remains unclear, since the incorporation of physical priors beyond the necessary constraints has not been well studied in this context.
The FQH setting differs because its strong correlations exhibit well-established patterns, such as the Laughlin wavefunction~\cite{laughlin_anomalous_1983}, composite fermions~\cite{jain_composite-fermion_1989}, and partons~\cite{jain_incompressible_1989}.
Such knowledge can be supplied to the network from outside the ansatz, as in the supervised pretraining on analytical wavefunctions performed by Nazaryan et al.~\cite{nazaryan_needle_2025}, which helps the network converge in subsequent variational energy optimizations.
Alternatively, it can be built into the ansatz itself, and whether this second route degrades the flexibility of the network, either by limiting its expressivity or by making it more difficult to optimize, remains largely unexplored.
Striking this balance is difficult, because encoding too little physical structure can render the relevant correlations difficult to learn, whereas rigidly imposing an approximate pattern can overconstrain the network.

In this work, we demonstrate that additional physical priors can be encoded without compromising the flexibility of the network, provided that these priors are not only relaxed but also embedded in structures that give the network the flexibility to improve upon them.
We first analyze how the construction of the monopole harmonic product orbital (MHPO)~\cite{qian_deephall_2025} ansatz makes it difficult to optimize and then compare it with a fixed Laughlin extension, which, while useful, constrains the network so strongly that its expressivity is impaired.
These two extremes motivate our balanced design, the orbital pair matrix factor MHPO (MHPO-M), which dresses each generalized orbital with a learnable, orbital-dependent pair factor: a multiplicative product that explicitly couples the electron at which the orbital is evaluated to every other electron.
In the tested $\nu=1/5$ and $1/3$ systems, as well as the non-Laughlin $\nu=2/5$ system, MHPO-M achieves lower variational energies than both bare MHPO and fixed Laughlin pair factors while avoiding the incorrect short-range physics imposed by the latter.
Because this balanced design frees the orbital readout from carrying the full correlation structure, we further propose a factorized orbital readout that extends MHPO-M to 35 electrons, yielding energies competitive with large-system estimates based on three-LL ED.

\section{Models and Methods}

\subsection{Spherical geometry}

Following Ref.~\cite{qian_deephall_2025}, we focus on spherical geometry~\cite{haldane_fractional_1983}, in which electrons are confined to a spherical surface, forming a 2D electron gas.
A magnetic monopole placed at the center of the sphere produces a total flux of $2Q\phi_0$ through the surface, where $Q$ is the monopole strength, $2Q$ is an integer, and $\phi_0$ is the flux quantum.
For simplicity, we assume $Q>0$ throughout.
On the sphere, the flux and the filling factor $\nu$ are related by $2Q=N/\nu-\mathcal{S}$, where $\mathcal{S}$ is the shift~\cite{wen_shift_1992}.
The systems studied here ($\nu=1/5$, $2/5$, and $1/3$) have $\mathcal{S}=5$, $4$, and $3$, respectively.
The sphere radius is chosen as $R=\sqrt{Q}\ell$, where $\ell = \sqrt{\hbar c/(eB)}$ is the magnetic length and $B$ is the uniform magnetic field strength on the sphere.
In the symmetric gauge $\mathbf{A}=\hbar Q/e R$, the many-body Hamiltonian is
\begin{equation}\label{eq:H}
    \hat H = \hbar \omega_c \left[ \sum_j\frac{\ell^2}{2R^2}|\hat{\boldsymbol{\Lambda}}_j|^2 + \sum_{j<k}\frac{\kappa\ell}{|\mathbf{r}_j-\mathbf{r}_k|}\right],
\end{equation}
where $j$ and $k$ label electrons, and
$\kappa \equiv \frac{e^2/(\epsilon\ell)}{\hbar\omega_c}$ is the LL mixing strength parameter, with $\epsilon$ the dielectric constant of the material.
Unless otherwise stated, all calculations use $\kappa=1$.
$\omega_c=eB/(m_b c)$ is the cyclotron frequency, $m_b$ is the band mass of the electrons, and $\hat{\boldsymbol{\Lambda}}_j$ is proportional to the canonical momentum tangential to the surface:
\begin{equation}\label{eq:lambda}
    |\hat{\boldsymbol{\Lambda}}_j|^2=-\frac{1}{\sin \theta_j}\frac{\partial}{\partial \theta_j}\sin\theta_j\frac{\partial}{\partial \theta_j}
    +\left(Q\cot\theta_j+\frac{\mathrm{i}}{\sin \theta_j}\frac{\partial}{\partial \phi_j}\right)^2.
\end{equation}

\subsection{Exact constraints for neural networks}\label{sec:principles}

In spherical geometry, obtaining meaningful variational energies requires the neural network wavefunction to satisfy exact physical constraints, including electron permutation antisymmetry and the correct asymptotic behavior around the poles.

Neglecting spin degrees of freedom, the real-space wavefunction ansatz $\psi_T$ must be antisymmetric under electron permutations.
This requirement is commonly fulfilled by taking the product of a permutation-antisymmetric determinant and a permutation-symmetric Jastrow factor $\mathrm{e}^J$~\cite{pfau_ferminet_2020, glehn_psiformer_2023, qian_deephall_2025}:
\begin{equation}\label{eq:ansatz}
    \psi_T(\mathbf{r}_1,\dots,\mathbf{r}_N) = \mathrm{e}^{J(\mathbf{r}_1,\dots,\mathbf{r}_N)}\sum_d \det\left[\varphi_{di} (\mathbf{r}_j;\{\mathbf{r}_{\ne j}\})\right].
\end{equation}
Here, $N$ is the number of electrons, $J$ is a permutation-symmetric scalar function, $d$ runs over the selected determinants, $i=1,\ldots,N$ indexes generalized orbitals (the determinant rows), and $j=1,\ldots,N$ indexes electrons (the determinant columns).
The function $\varphi_{di}(\mathbf{r}_j;\{\mathbf{r}_{\ne j}\})$ denotes a generalized orbital, with $\mathbf{r}_j$ representing the coordinate of the $j$th electron.
The notation $\{\mathbf{r}_{\ne j}\}$ indicates that the orbital's dependence on the coordinates of other electrons remains invariant under their permutations, thereby maintaining the overall antisymmetry of the wavefunction.
In this work, $J$ is chosen to have the following form to satisfy the two-dimensional Coulomb cusp condition~\cite{kato_cusp_1957}
\begin{equation}\label{eq:jastrow}
    J(\mathbf{r}_1,\dots,\mathbf{r}_N) = -\frac{\kappa}{3}\sum_{j<k}\frac{r_0^2}{r_0 + |\mathbf{r}_j-\mathbf{r}_k|},
\end{equation}
where $r_0$ is a learnable scalar initialized to one.

Beyond permutation antisymmetry, the spherical geometry imposes additional requirements on the wavefunction.
Since the electron density near the poles remains nonzero, if the terms in the parentheses of Eq.~\eqref{eq:lambda} fail to cancel exactly for electron $j$, the kinetic contribution to $\hat H\psi_T$ will diverge near the poles as $1/\sin^2 \theta_j$, leading to a divergent total energy and thus unstable optimization.
Therefore, for each electron $j$ with all other electron coordinates fixed, the following singularity condition at the poles must be imposed when designing the network:
\begin{equation}\label{eq:singularity-condition}
    \psi_T(\mathbf{r}_1,\dots,\mathbf{r}_N) \propto \begin{cases}
        \exp (\mathrm{i} Q \phi_j) & \theta_j \to 0,\\
        \exp (-\mathrm{i} Q \phi_j) & \theta_j \to \pi.
    \end{cases}
\end{equation}
The network designs are described in detail in Secs.~\ref{sec:mhpo}--\ref{sec:mhpo-m}.

\subsection{Computational details}

Subject to these exact constraints, we parametrize the wavefunction using a neural network based on the Psiformer~\cite{glehn_psiformer_2023} architecture, following Ref.~\cite{qian_deephall_2025}.
The resulting per-electron feature channels are denoted by $f_\alpha(\mathbf{r}_j;\{\mathbf{r}_{\ne j}\})$ in the ansatzes introduced below.
The energy expectation value of the trial wavefunction is estimated using Markov chain Monte Carlo, and the parameters are optimized along the energy descent direction with Kronecker-factored approximate curvature (K-FAC)~\cite{martens_kfac_2015}.
The code is implemented within the JaQMC framework~\cite{ren_towards_2023, bytedance_jaqmc}, and the network and training hyperparameters are summarized in the Supplemental Material.

For comparison, we perform ED calculations in the sector with vanishing total angular momentum projection along the monopole axis, $L_z=0$, retaining up to the lowest three Landau levels.
Unless otherwise stated, the corresponding Hilbert spaces are untruncated.
Where an excitation-number cutoff is required, configurations in which more than the specified number of electrons are excited above the lowest Landau level are discarded.

\subsection{AI Usage}

The ED code was developed with assistance from ByteDance internal AI tools.
Our ED implementation was validated against the DiagHam library~\cite{diagham}.
The code for data analysis, figure plotting, and the network ansatz was also written with assistance from ByteDance internal AI tools under author-specified prompts, and was reviewed, modified as needed, and independently verified by the authors before use in producing the reported results.

\begin{figure}
    \centering
    \includegraphics{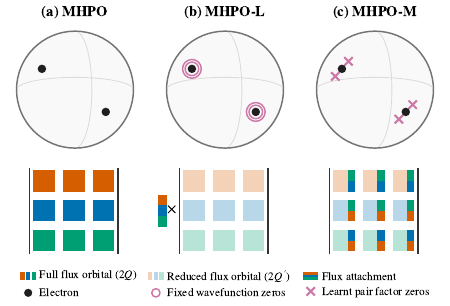}
    \caption{%
        Schematic illustration of the correlation-zero priors and the monopole-harmonic degree allocation in the three neural wavefunction constructions.
        \textbf{(a)} In MHPO [Eq.~\eqref{eq:mhpo-orbital}], the zeros at electron coincidences are fixed by the determinant structure, while the correlation zeros are left entirely for the network to learn, and the monopole harmonics in the orbitals are constructed at the physical flux $2Q$.
        \textbf{(b)} In the Laughlin wavefunction [Eq.~\eqref{eq:laughlin-det}] and in the Laughlin-flux MHPO (MHPO-L) ansatz it inspires [Eq.~\eqref{eq:mhpo-l-orbital}], the Laughlin flux attachment Jastrow factor imposes additional correlation zeros of order $2p$ at every electron coincidence, and the orbitals are constructed at the reduced effective flux $2Q^*=2Q-2p(N-1)$.
        Here, $p$ is a positive integer called the flux-pair number, so that $2p$ flux quanta are attached to each electron, while the remaining monopole-harmonic degree is supplied by the global Laughlin factor $\prod_{j<k}(u_j v_k-u_k v_j)^{2p}$.
        We use $p=2$ for $\nu=1/5$ and $p=1$ for $\nu=1/3$ and $2/5$.
        \textbf{(c)} In the orbital pair matrix factor MHPO (MHPO-M) ansatz [Eq.~\eqref{eq:mhpo-m-orbital}], the same reduced-flux orbitals are dressed by orbital- and configuration-dependent pair factors, which are initialized in the Laughlin form but remain optimizable.
        These pair factors, together with the orbitals, jointly determine the correlation structure of the whole wavefunction, thereby making the zero structure both physically informed and learnable.
    }
    \label{fig:schematic}
\end{figure}

\begin{figure*}
    \centering
    \includegraphics{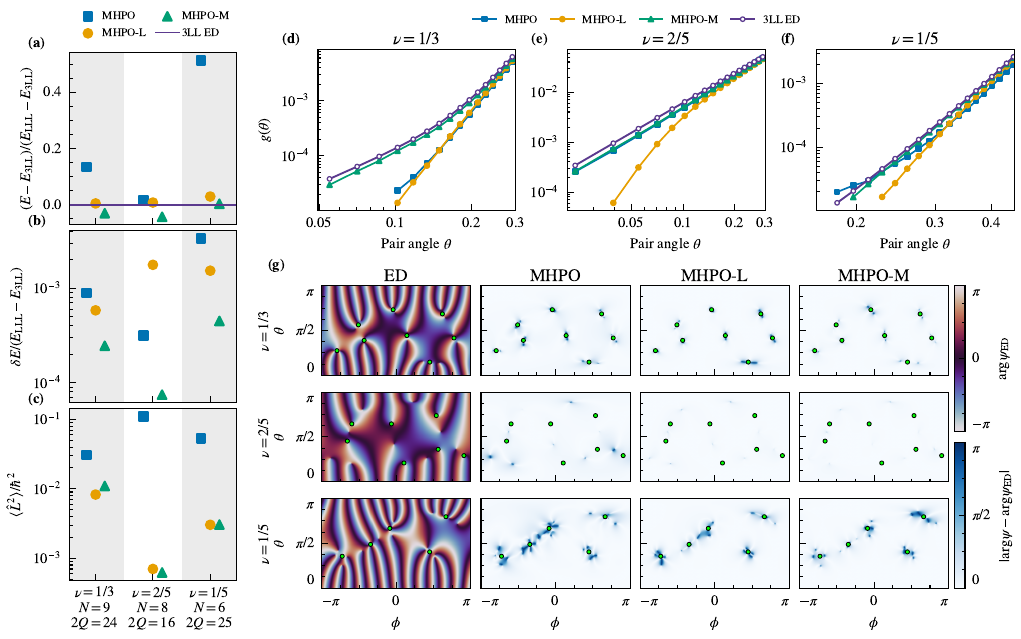}
    \caption{%
        Comparison of the MHPO (blue), MHPO-L (yellow), and MHPO-M (green) ansatzes for the $\nu=1/3$ ($N=9$, $2Q=24$), $\nu=2/5$ ($N=8$, $2Q=16$), and $\nu=1/5$ ($N=6$, $2Q=25$) systems.
        \textbf{(a)} Normalized and offset variational energy $(E-E_{\text{3LL}})/(E_{\text{LLL}}-E_{\text{3LL}})$, where $E_{\text{LLL}}$ and $E_{\text{3LL}}$ are exact-diagonalization energies obtained in the LLL and three-LL Hilbert spaces, so that values of one and zero correspond to these two references.
        \textbf{(b)} Corresponding autocorrelation-corrected statistical error of the mean energy over 10000 evaluation steps, normalized by $(E_{\text{LLL}}-E_{\text{3LL}})$.
        \textbf{(c)} Corresponding total-angular-momentum expectation values $\langle \hat L^2\rangle/\hbar^2$, shown on a logarithmic scale.
        \textbf{(d--f)} Short-range pair correlation function $g(\theta)$ for the $\nu=1/3$, $2/5$, and $1/5$ systems, with three-LL ED shown in violet.
        Points with tiny PCF values, for which the relative noise is large, are removed.
        \textbf{(g)} Phase comparison, with the three systems shown in separate rows.
        The first column gives the three-LL ED phase $\operatorname{arg}\psi_\text{ED}$, while the remaining columns give the absolute value of the wrapped phase difference $|\operatorname{arg}\psi-\operatorname{arg}\psi_\text{ED}|$ for MHPO, MHPO-L, and MHPO-M.
        All phases and phase differences are wrapped to $[-\pi,\pi]$, and green circles mark the fixed electron positions.
        The ED calculations for the $\nu=1/5$ and $2/5$ systems are untruncated, whereas those for the $\nu=1/3$ system are restricted to configurations with at most three electrons excited above the LLL.
        In (a) and (c), the statistical error bars are smaller than the markers.
        All error bars in this work reflect Monte Carlo sampling uncertainty only and do not account for stochasticity in the optimization.
    }
    \label{fig:energy-phases}
\end{figure*}

\section{Results}

Although the exact constraints in Sec.~\ref{sec:principles} ensure that the wavefunction is physically plausible, they leave open how electron correlations should be represented within the ansatz.
We therefore examine different ways of encoding additional physical priors in Secs.~\ref{sec:mhpo}--\ref{sec:mhpo-m}.

\subsection{Monopole harmonic product orbital ansatz}\label{sec:mhpo}

The monopole harmonic product orbital (MHPO) ansatz introduced in Ref.~\cite{qian_deephall_2025} satisfies the exact physical constraints of permutation antisymmetry and the singularity condition in Eq.~\eqref{eq:singularity-condition}, while leaving the correlations to be handled entirely by the neural network.
It takes the same form as Eq.~\eqref{eq:ansatz}, with the generalized orbitals given by
\begin{multline}\label{eq:mhpo-orbital}    
    \varphi^\text{MHPO}_{Q;di}(\mathbf{r}_j;\{\mathbf{r}_{\ne j}\}) = \sum_{\alpha ,m} P^\text{MHPO}_{di\alpha m} \times \\f_{\alpha}(\mathbf{r}_j;\{\mathbf{r}_{\ne j}\}) u_j^{Q+m} v_j^{Q-m},    
\end{multline}    
where $d$ runs over the selected determinants, $i=1,\dots,N$ indexes the generalized orbitals, and $P^\text{MHPO}_{di\alpha m}$ are complex parameters.
The permutation-invariant neural network features $f_{\alpha}(\mathbf{r}_j;\{\mathbf{r}_{\ne j}\})$, with $\alpha$ indexing the selected feature channels, carry the electron correlations.
The factors $u_j^{Q+m} v_j^{Q-m}$, with $m=-Q,-Q+1,\dots,Q$, are monopole harmonics~\cite{wu_dirac_1976} and form the eigenwavefunctions of the LLL.
They are built from the spinor coordinates of the $j$-th electron, $u_j=\cos(\theta_j/2) \mathrm{e}^{\mathrm{i} \phi_j/2}$ and $v_j=\sin(\theta_j/2) \mathrm{e}^{-\mathrm{i} \phi_j/2}$.
Each monopole harmonic is a homogeneous monomial of degree $2Q$ in these spinor coordinates, and we refer to this degree as the monopole-harmonic degree.
This allocation of monopole-harmonic degrees of freedom is illustrated in Fig.~\ref{fig:schematic}(a).
While the determinant structure naturally ensures zeros when electrons coincide, it does not explicitly encode the additional correlation zeros, leaving them to be represented implicitly through the learned neural features.
Because every monopole-harmonic term carries the same monopole-harmonic degree $2Q$ for electron $j$ and the learned features do not alter this degree, the ansatz enforces the required pole behavior.
The monopole-harmonic construction also allows the neural network to be initialized near the LLL, enabling efficient optimization.
Because the monopole harmonics carry the gauge dependence explicitly, the network only needs to learn gauge-invariant correlations, which further eases optimization.

However, the MHPO ansatz is not generally guaranteed to provide optimal flexibility.
In particular, whether the MHPO construction can exactly represent the Laughlin wavefunction, given unlimited neural network expressivity for regular functions, is analytically nontrivial.
Even if such a representation exists, reproducing the clusters of $2p$ zeros associated with each electron pair at filling $\nu=1/(2p+1)$ for larger $N$ would require coefficient functions of great complexity.
A finite network must represent and optimize these functions from scratch, making the relevant correlations difficult to learn in practice, consistent with the observed behavior of MHPO at lower fillings.

Figure~\ref{fig:energy-phases}(a) shows that, for LL mixing parameter $\kappa=1$, the MHPO network achieves energies lower than LLL ED but higher than three-LL ED for all systems studied, where three-LL ED corresponds to the largest number of LLs accessible to our ED calculations.
On the normalized vertical axis, energies lying between these two references therefore fall between 0 and 1.
Moreover, the converged energy for the $\nu=1/5$ system with 6 electrons is significantly worse than that of the $\nu=2/5$ system with 8 electrons and the $\nu=1/3$ system with 9 electrons, despite the similar total flux ($2Q=25$ for $\nu=1/5$ and $2Q=24$ for $\nu=1/3$).
The phase comparison in Fig.~\ref{fig:energy-phases}(g) shows the three-LL ED phase obtained by fixing all electrons but one, together with the differences between the network wavefunction phases and the three-LL ED phase.
The ED phase at $\nu=1/5$ exhibits more zeros, and these zeros are more strongly clustered.
At this filling, MHPO also deviates substantially more from the three-LL ED phase than at the other two fillings, providing a complementary view of the limitation discussed above.

If the MHPO construction allowed a different allocation of monopole-harmonic degrees across orbitals, learning the Laughlin wavefunction would become considerably easier.
Up to normalization, the Laughlin wavefunction at filling $\nu=1/(2p+1)$ can be written as
\begin{equation}\label{eq:laughlin-det}  
    \psi_\text{Laughlin} \propto \det_{m,j} \left[ u_j^{Q^*+m}v_j^{Q^*-m} \mathcal{J}^\text{Laughlin}_j \right],  
\end{equation}  
where $p$ is a positive integer so that $2p$ flux quanta are attached to each electron, $Q^*=Q-p(N-1)$, $m=-Q^*,\ldots,Q^*$ labels the effective-LLL orbitals, and explicit electron correlations enter through the fixed orbital pair factor $\mathcal{J}^\text{Laughlin}_j = \prod_{k \ne j} (u_j v_k - u_k v_j)^p$.
Here, each orbital carries monopole-harmonic degrees of freedom from all electrons, which the MHPO construction cannot achieve on an orbital-by-orbital basis.

\subsection{Fixed Laughlin pair factors}\label{sec:mhpo-l}

Given the limitations of the MHPO ansatz, a natural extension is to incorporate the correlation structure of the Laughlin wavefunction~\cite{laughlin_anomalous_1983, jain_composite_2007}, yielding the Laughlin-flux MHPO (MHPO-L) ansatz:
\begin{multline}\label{eq:mhpo-l-orbital}  
    \varphi^\text{MHPO-L}_{Q;di}(\mathbf{r}_j;\{\mathbf{r}_{\ne j}\}) = \varphi^\text{MHPO}_{Q^*;di}(\mathbf{r}_j;\{\mathbf{r}_{\ne j}\}) \mathcal{J}^\text{Laughlin}_j.  
\end{multline}  
Here, the orbital sector is constructed at the reduced effective flux $2Q^*=2Q-2p(N-1)$.
To test whether a Laughlin-inspired prior can also benefit non-Laughlin states, we apply the same flux-attachment factor not only at $\nu=1/3$ and $1/5$, where the Laughlin wavefunction provides a direct starting point, but also at $\nu=2/5$, where the Laughlin wavefunction itself is inapplicable and the natural composite-fermion prior is instead the Jain wavefunction~\cite{jain_composite-fermion_1989, jain_quantitative_1997}.
The Jain wavefunction, however, requires projection onto the LLL, which is not directly compatible with a real-space ansatz, whereas the flux-attachment factor $\mathcal{J}^\text{Laughlin}_j$ is well-defined at any filling and requires no projection, making it a natural real-space prior to explore even outside the Laughlin sequence.
As illustrated in Fig.~\ref{fig:schematic}(b), the MHPO-L ansatz imposes, beyond the zero already enforced by the determinant, an additional zero of order $2p$ at every electron coincidence.
Relative to the bare MHPO ansatz, which contains no explicit pair-factor structure, this fixed Laughlin pair factor improves both the variational energy and the angular momentum across all three fillings, and it reduces the autocorrelation-corrected statistical error at $\nu=1/3$ and $1/5$ while increasing it substantially at $\nu=2/5$, as shown in Fig.~\ref{fig:energy-phases}(a--c).

The increased statistical error at $\nu=2/5$ is an early symptom of a deeper problem with the fixed pair factor.
Although the $2p$ additional Laughlin zeros are mathematically available, they give an incorrect picture at $\nu=2/5$ and are not exact even at fillings $\nu=1/(2p+1)$.
Moreover, this construction promotes the approximate Laughlin pattern to a hard structural constraint, leaving the network little room to adjust it.
In reality, the Coulomb ground state generally retains a small but nonzero pair amplitude in the channel with relative angular momentum equal to 1, and LL mixing relaxes this constraint further.
As illustrated in Fig.~\ref{fig:energy-phases}(g), the correlation zeros in the $\nu=1/3$ and $1/5$ systems remain close to, but do not exactly coincide with, the electrons, whereas in the $\nu=2/5$ system they are distributed approximately at the midpoints between neighboring electron pairs.
Surprisingly, at the level of the global phase, MHPO-L agrees more closely with the three-LL ED phase than MHPO does at $\nu=2/5$, indicating that the neural network cancels the excess $4\pi$ phase winding around the electrons by generating anti-vortices at the electron positions and develops new correlation zeros at the correct locations.
The fixed Laughlin zeros themselves, however, cannot be removed.
The consequences are visible in the short-range pair correlation function (PCF) $g(\theta)$, shown on a logarithmic scale in Fig.~\ref{fig:energy-phases}(d--f), where $\theta$ denotes the scalar angular separation between two electrons.
Across all three fillings, MHPO-L exhibits a steeper short-range decay than three-LL ED, causing $g(\theta)$ to vanish too rapidly as $\theta\rightarrow0$.

\begin{figure*}
    \centering
    \includegraphics{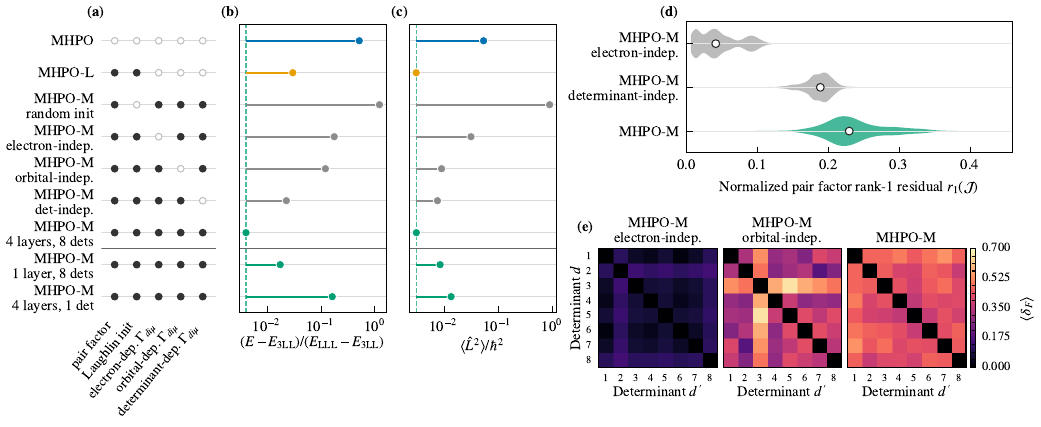}
    \caption{%
\textbf{(a)} Network settings used in the ablation study.
Filled circles indicate enabled properties.
``Pair factor'' denotes the application of explicit factors $\mathcal{J}$; when this is the only enabled property, $\mathcal{J}$ corresponds to the fixed Laughlin factor $\mathcal{J}^\text{Laughlin}$.
``Laughlin init'' indicates that $\Gamma_{di\mu}$ is initialized according to Eq.~\eqref{eq:laughlin-init}.
``electron-dependent $\Gamma_{di\mu}$'' makes $\Gamma_{di\mu}$ depend on the neural network features according to Eq.~\eqref{eq:gamma-from-network}, and hence on the electron configuration, whereas disabling this property reduces $\Gamma_{di\mu}$ to a plain parameter to be optimized.
``Orbital-dependent $\Gamma_{di\mu}$'' and ``Determinant-dependent $\Gamma_{di\mu}$'' assign a distinct $\Gamma_{di\mu}$ to each orbital or to each determinant, respectively.
The default network construction uses 4 layers and 8 determinants.
Rows below the horizontal line retain the full MHPO-M construction and vary only the network capacity, namely the number of neural network layers and determinants, with each reduced to one.
\textbf{(b)} Normalized and offset variational energy $(E-E_{\text{3LL}})/(E_{\text{LLL}}-E_{\text{3LL}})$ for the different network settings.
Here, $E_{\text{LLL}}$ and $E_{\text{3LL}}$ are exact diagonalization energies obtained in the LLL and three-LL Hilbert spaces, respectively.
\textbf{(c)} Corresponding dimensionless total-angular-momentum expectation values $\langle \hat L^2\rangle/\hbar^2$, shown on a logarithmic scale.
\textbf{(d)} Normalized rank-1 residual $r_1(\boldsymbol{\mathcal{J}}_d)$ of the orbital-by-electron pair-factor matrix for the orbital-dependent $\Gamma_{di\mu}$ settings.
\textbf{(e)} Average normalized Frobenius distance $\langle \delta F_{dd'} \rangle$ between the pair-factor matrices $\boldsymbol{\mathcal{J}}_d$ and $\boldsymbol{\mathcal{J}}_{d'}$ of different wavefunction determinants for the determinant-dependent $\Gamma_{di\mu}$ settings.
All results in this figure are obtained for the $\nu=1/5$ ($N=6$, $2Q=25$) system.
In (b--c), the statistical error bars are smaller than the markers.
}
    \label{fig:ablation}
\end{figure*}
\begin{figure*}
    \centering
    \includegraphics{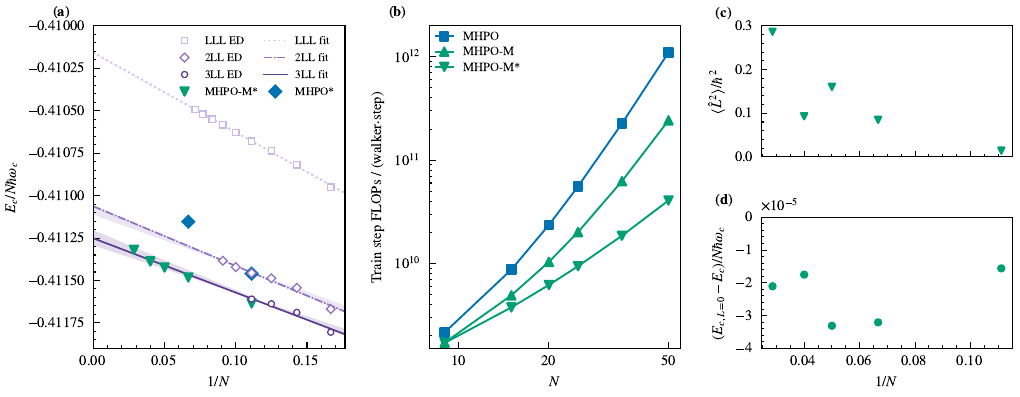}
    \caption{%
\textbf{(a)} Corrected converged energies $E_c$ of MHPO* and MHPO-M* wavefunctions for $\nu=1/3$ systems with $N=9$ to $35$ electrons.
The corrected energy includes the neutralizing background-charge contribution, subtracts the kinetic energy $N\hbar\omega_c/2$, and incorporates the density correction that suppresses the system-size dependence~\cite{morf_microscopic_1986, jain_composite_2007, qian_deephall_2025}.
The LLL, two-LL, and three-LL ED energies are shown as violet markers, and the dashed lines are $1/N$ fits.
The shaded bands show the change caused by omitting the smallest system.
For the two-LL and three-LL results, the shading is extended to reflect the alternative trend inferred by fitting the four available energies directly.
The two-LL ED energies for the $N=10$ and $11$ systems are truncated to include at most two electrons excited above the LLL, and the three-LL ED energy for the $N=9$ system is truncated to include at most three electrons excited above the LLL, while the remaining systems are untruncated.
\textbf{(b)} Floating-Point Operations (FLOPs) per training step per walker for the MHPO, MHPO-M, and MHPO-M* ansatzes at different electron numbers, excluding Monte Carlo sampling FLOPs between steps.
\textbf{(c)} Converged dimensionless total-angular-momentum expectation values $\langle \hat L^2\rangle/\hbar^2$ of the same MHPO-M* wavefunctions as in (a).
\textbf{(d)} Difference between the corrected energies and the corrected energies projected onto the $L=0$ sector for the MHPO-M* wavefunctions.
The statistical error bars are smaller than the markers in (a), (c), and (d).
    }
    \label{fig:scaling}
\end{figure*}

\subsection{Learnable orbital pair factors}\label{sec:mhpo-m}

Motivated by these observations, we propose the orbital pair matrix factor MHPO (MHPO-M) ansatz, which retains the monopole-harmonic degree allocation of MHPO-L while relaxing its predetermined Laughlin zeros.
This keeps the explicit pair-factor structure that the bare MHPO ansatz lacks, without fixing it so rigidly that the network cannot adjust it, thereby balancing the physical prior against the flexibility of the network:
\begin{gather}  \label{eq:mhpo-m-orbital}
    \varphi^\text{MHPO-M}_{Q;di}(\mathbf{r}_j;\{\mathbf{r}_{\ne j}\}) = \varphi^\text{MHPO}_{Q^*;di}(\mathbf{r}_j;\{\mathbf{r}_{\ne j}\}) \times \mathcal{J}_{dij},\\  
    \mathcal{J}_{dij}=\prod_{k\ne j} \prod_{\mu=1}^p \begin{pmatrix}  
        u_j & v_j  
    \end{pmatrix}  
    \Gamma_{di\mu}(\mathbf{r}_j;\{\mathbf{r}_{\ne j}\})  
    \begin{pmatrix}  
        u_k\\ v_k  
    \end{pmatrix},    \label{eq:j-from-gamma}  
\end{gather}  
Here, the scalar $\mathcal{J}_{dij}$ is the orbital pair factor dressing orbital $i$ in determinant $d$ when it is evaluated at electron $j$.
It explicitly couples the spinor coordinates of electron $j$ to those of every other electron $k$, with $\mu=1,\ldots,p$ labeling the channels associated with the $p$ attached flux pairs.
The $2\times 2$ matrix $\Gamma_{di\mu}$ depends on the orbital and on the electron configuration through the features $f_{\alpha}$:
\begin{equation}  \label{eq:gamma-from-network}
    \Gamma_{di\mu}(\mathbf{r}_j;\{\mathbf{r}_{\ne j}\}) = \sum_\alpha W_{\alpha di\mu} f_{\alpha}(\mathbf{r}_j;\{\mathbf{r}_{\ne j}\}) + B_{di\mu}.  
\end{equation}  
Here, $W_{\alpha di\mu}$ and $B_{di\mu}$ are learnable $2\times 2$ matrices.
Through the dependence of $\Gamma_{di\mu}$ on the neural features, $\mathcal{J}_{dij}$ is a function of the full electron configuration.
The Laughlin pair factor is recovered by choosing
\begin{equation}  \label{eq:laughlin-init}
    \Gamma_{di\mu}=\begin{pmatrix}0 & 1\\ -1 & 0\end{pmatrix},\  \forall d,i,\mu.
\end{equation} 
We adopt this choice to initialize the learnable variants of the matrix-factor construction, so that the optimization begins from an explicit, physically motivated structure.

As illustrated in Fig.~\ref{fig:schematic}(c), the MHPO-M ansatz employs a monopole-harmonic degree allocation scheme similar to that of the MHPO-L ansatz.
However, because the matrices can become orbital-, determinant-, and electron-configuration-dependent during optimization, the resulting pair factors generally cannot be extracted as a common fixed factor of the many-body wavefunction.
The zeros can therefore move or relax in response to the variational objective, and the wavefunction zeros and the angular momentum are jointly determined by the pair factors and the neural network.

As demonstrated in Fig.~\ref{fig:energy-phases}(a--b), the MHPO-M ansatz achieves lower variational energies and smaller statistical errors than both MHPO and MHPO-L across all tested fillings.
At $\nu=2/5$, the rigid Laughlin prior in MHPO-L yields only a modest energy gain over MHPO at the cost of a much larger statistical error, whereas MHPO-M substantially improves both metrics.
Even though the Laughlin wavefunction does not itself describe the $\nu=2/5$ state, combining its flux-attachment prior with learnable pair factors enables the network to reshape the mismatched correlation structure into an accurate ground state.
Furthermore, the total angular momentum of MHPO-M is consistently lower than that of MHPO, while its performance relative to MHPO-L varies across fillings [Fig.~\ref{fig:energy-phases}(c)].
Because the fixed Laughlin pair factors $\mathcal{J}^\text{Laughlin}_j$ do not couple to the neural network and contribute exactly zero total angular momentum, MHPO-L can retain an advantage over MHPO-M in angular momentum even when its variational energy is higher.

The synergy between the physical prior and network flexibility is even clearer in the phase structure and the short-range correlations.
Figure~\ref{fig:energy-phases}(g) compares the phase structures of the three ansatzes with three-LL ED.
At $\nu=1/3$, all three ansatzes exhibit similar phase accuracy.
At $\nu=2/5$ and $1/5$, MHPO shows larger deviations, whereas MHPO-L and MHPO-M remain much closer to the ED phase.
Although the residual phase difference is larger for the $\nu=1/5$ system than for the $\nu=2/5$ system, it is concentrated near three closely spaced zeros, where the wavefunction amplitude is small.
In the short-range PCFs shown in Fig.~\ref{fig:energy-phases}(d--f), MHPO-M consistently follows the three-LL ED results across all three fillings, eliminating the artificial short-range suppression imposed by the fixed Laughlin zeros, most notably at $\nu=2/5$.

To identify which ingredients of the prior are required for optimal flexibility, we compare constrained variants of the MHPO-M ansatz for the $\nu=1/5$ system with $N=6$ electrons.
These variants separately disable the Laughlin initialization [Eq.~\eqref{eq:laughlin-init}] and the electron-configuration, orbital, or determinant dependence of the $\Gamma_{di\mu}$ matrix in Eq.~\eqref{eq:j-from-gamma} [Fig.~\ref{fig:ablation}(a)].
Throughout this family, we write $\Gamma_{di\mu}$ and $\mathcal{J}_{dij}$ generically, omitting the electron coordinate arguments for brevity.
When $\Gamma_{di\mu}$ is fixed, we set it to the Laughlin value of Eq.~\eqref{eq:laughlin-init}, so that the construction reduces to MHPO-L.

As shown in Fig.~\ref{fig:ablation}(a--c), dropping the Laughlin initialization [Eq.~\eqref{eq:laughlin-init}] has the most notable impact, and the result converges to a higher energy and angular momentum than even bare MHPO.
Proper initialization is therefore crucial for the optimization.
Sharing $\Gamma_{di\mu}$ across orbitals (orbital-independent MHPO-M) or making it independent of the electron configuration (electron-independent MHPO-M) degrades the results to a comparable extent.
Although these variants relax the Laughlin constraint and exhibit better expressivity than MHPO-L, they fail to optimize to a lower energy or angular momentum than MHPO-L.
This indicates that simply relaxing the Laughlin constraint does not necessarily improve flexibility.
The determinant dependence of $\Gamma_{di\mu}$, while less critical than the previous two ingredients, also plays an important role.
Without it (determinant-independent MHPO-M), the energy is only slightly better than that of MHPO-L, and the angular momentum remains larger.
Only when the electron-configuration, orbital, and determinant dependence of $\Gamma_{di\mu}$ are all included, forming the full MHPO-M ansatz, are the lowest energy and the smallest angular momentum contamination achieved.

To test whether the learned pair factors develop nonseparable orbital--electron dependence, we collect them into an $N\times N$ matrix $\boldsymbol{\mathcal{J}}_d=[\mathcal{J}_{dij}]_{i,j=1}^N$ per determinant and measure the normalized rank-1 residual
\begin{equation}
    r_1(\boldsymbol{\mathcal{J}}_d)=\min_{\operatorname{rank}(\mathbf{X})=1} \frac{\|\boldsymbol{\mathcal{J}}_d-\mathbf{X}\|_F}{\|\boldsymbol{\mathcal{J}}_d\|_F},
\end{equation}
which vanishes when $\boldsymbol{\mathcal{J}}_d$ separates into independent orbital and electron contributions.
All three orbital-dependent settings exhibit a statistically significant nonzero residual [Fig.~\ref{fig:ablation}(d)], indicating that they deviate from a simple Laughlin factor, with the full MHPO-M displaying the largest residual and hence the strongest nonseparable dependence.

We next examine how network depth and determinant count affect the performance of MHPO-M.
Reducing the network to a single layer still yields energies comparable to MHPO-L and far better than bare MHPO, whereas reducing to a single determinant degrades the energy much more, indicating that determinant count can matter more than network depth [Fig.~\ref{fig:ablation}(a--c)].
The robustness to reduced network depth suggests that the MHPO-M construction makes the reduced-flux orbitals much easier to learn.
However, unlike the Laughlin factor, MHPO-M does not preserve angular momentum, so the neural network must still incorporate many-body correlations and drive the angular momentum toward zero.
To quantify how distinct the pair factors are across determinants, we define the normalized Frobenius distance
\begin{equation}
    \delta F_{dd'}=\left\|\frac{\boldsymbol{\mathcal{J}}_d}{\|\boldsymbol{\mathcal{J}}_d\|_F}-\frac{\boldsymbol{\mathcal{J}}_{d'}}{\|\boldsymbol{\mathcal{J}}_{d'}\|_F}\right\|_F.
\end{equation}
The Monte Carlo averages $\langle \delta F_{dd'} \rangle$ [Fig.~\ref{fig:ablation}(e)] follow a similar trend, with electron-independent MHPO-M pair factors nearly identical across determinants, whereas full MHPO-M pair factors differ strongly, demonstrating that the multideterminant models exploit their flexibility to realize distinct pair factors.

\subsection{Scalable factorization for larger systems}

The preceding results established the accuracy benefits of the balanced MHPO-M construction.
This construction also offers a second advantage, namely efficiency.
The computational bottleneck lies in the MHPO readout of Eq.~\eqref{eq:mhpo-orbital}, which contains a weight tensor $P^\text{MHPO}_{di\alpha m}$ whose size scales quadratically with the electron number $N$ at fixed filling.
Because MHPO-M transfers part of the correlation structure into explicit pair factors, the remaining reduced-flux orbital readout can be simplified, improving the memory and computational scaling with $N$ and enabling access to larger systems.
To this end, we write the factorized MHPO (MHPO*) orbitals as
\begin{multline}\label{eq:factorized-orbital}  
    \varphi^\text{MHPO*}_{Q;di}(\mathbf{r}_j;\{\mathbf{r}_{\ne j}\}) = \sum_{\alpha ,m} P^\text{MHPO*}_{di\alpha m} \times \\ f_{\alpha}(\mathbf{r}_j;\{\mathbf{r}_{\ne j}\}) u_j^{Q+m} v_j^{Q-m},  
\end{multline}  
where the weight tensor $P^\text{MHPO*}_{di\alpha m}$ is factorized into a learnable tensor $\mathcal{A}_{d\alpha m}$ and a learnable matrix $\mathcal{B}_{i\alpha}$:
\begin{equation}\label{eq:factorized-factor}  
    P^\text{MHPO*}_{di\alpha m} = \mathcal{A}_{d\alpha m} \mathcal{B}_{i\alpha}.  
\end{equation}
Each factor in Eq.~\eqref{eq:factorized-factor} then scales linearly with the system size, as in neural network wavefunctions for molecules~\cite{pfau_ferminet_2020, glehn_psiformer_2023}.
The factorized MHPO-M (MHPO-M*) ansatz is obtained by replacing the reduced-flux MHPO orbital in the MHPO-M construction with the MHPO* orbital:
\begin{equation}\label{eq:MHPO-M*}
    \varphi^\text{MHPO-M*}_{Q;di}(\mathbf{r}_j;\{\mathbf{r}_{\ne j}\})
    = \varphi^\text{MHPO*}_{Q^*;di}(\mathbf{r}_j;\{\mathbf{r}_{\ne j}\})\mathcal{J}_{dij}.
\end{equation}
The factorization is thus applied only to the orbital readout, while the MHPO-M pair factors $\mathcal{J}_{dij}$ retain the form defined in Eq.~\eqref{eq:j-from-gamma}.
In practice, contracting the two factors sequentially also avoids materializing the quadratic tensor $P^\text{MHPO*}_{di\alpha m}$.

Figure~\ref{fig:scaling}(a) shows the converged energies of the MHPO* and MHPO-M* wavefunctions at $\nu=1/3$ for systems with $N = 9$ to $35$ electrons.
For reference, we also include the LLL, two-LL, and three-LL ED energies, together with linear fits in $1/N$ as a guide.
The largest untruncated ED systems accessible to us are $N=14$ (LLL) and $N=9$ (two-LL).
We additionally include truncated results for $N=10$ and $11$ in two-LL ED and for $N=9$ in three-LL ED, since the energies are insensitive to this cutoff.
For example, in the $N=9$ three-LL ED calculation at $\kappa=1$, restricting the Hilbert space to at most two rather than three electrons excited above the LLL changes the energy by only about $10^{-6}\hbar\omega_c$ per electron.
Within the ED-accessible regime, MHPO-M* reaches a lower variational energy than three-LL ED at $N=9$, consistent with the original MHPO-M, indicating that the factorized construction does not compromise energy accuracy at this scale.
For systems larger than 15 electrons, ED is no longer directly available, so we instead compare the size dependence of the MHPO-M* energies.
From $N=15$ to $25$, these energies follow a smooth $1/N$ scaling trend that agrees favorably with the trend inferred from three-LL ED, while at $N=35$ they deviate slightly from this trend and lie marginally above the three-LL ED extrapolation.
These results support the accuracy of the MHPO-M* ansatz at larger system sizes.
In contrast, the MHPO* ansatz is consistent with two-LL ED at $N=9$ but lies considerably above the two-LL ED trend at $N=15$.
Alongside this energy behavior, the converged MHPO-M* angular-momentum expectation value $\langle \hat L^2\rangle/\hbar^2$ grows with the number of electrons, as shown in Fig.~\ref{fig:scaling}(c).
Projecting onto the $L=0$ sector, however, changes the corrected MHPO-M* energy by only about $10^{-5} \hbar \omega_c$ per electron, as shown in Fig.~\ref{fig:scaling}(d), with the projection procedure and the projected-energy estimator described in the Supplemental Material.
The growing angular momentum therefore does not indicate that the high-$L$ components carry appreciable energy.
Taken together, these results demonstrate the scalability and accuracy of the MHPO-M* wavefunction ansatz and its potential for studying larger systems.

\section{Discussion}\label{sec:discussion}

In summary, we have shown that physical priors for FQH systems beyond the necessary constraints can be incorporated into NNFQH states in a balanced way that preserves their flexibility.
The bare MHPO ansatz leaves the network to learn implicit and intricate zero structures, and consequently yields lower accuracy at lower fillings, whereas fixed Laughlin factors, although they improve upon bare MHPO, introduce predetermined zeros whose resulting bias the neural network cannot overcome.
MHPO-M overcomes both limitations by retaining the Laughlin form as an initialization while optimizing orbital- and configuration-dependent pair factors, yielding the lowest energies among the tested variants at $\nu=1/5$ and $1/3$, as well as at the non-Laughlin filling $\nu=2/5$.
This construction further enables a factorized orbital readout that extends the approach to 35 electrons, with energies comparable to large-system estimates based on three-LL ED.
Together, these results demonstrate that physical priors can be incorporated in a balanced, learnable form that preserves, rather than restricts, the flexibility of the ansatz.

In traditional FQH studies, other methods have been proposed to overcome the limitations of Laughlin flux attachment, one of which is the Jain-Kamilla projection~\cite{jain_quantitative_1997}.
The Jain-Kamilla wavefunction $\psi_\text{JK}$ can be written as
\begin{equation}  
    \psi_\text{JK}(\mathbf{r}_1,\dots,\mathbf{r}_N) = \det\left[\hat{\mathcal{P}}_\text{LLL} [\varphi_{i} (\mathbf{r}_j) \mathcal{J}^\text{Laughlin}_j]\right],  
\end{equation}  
where $\hat{\mathcal{P}}_\text{LLL}$ is the LLL projection operator and $\varphi_{i} (\mathbf{r}_j)$ are filled LL orbitals.
A side effect of this projection is that the flux attachment $\mathcal{J}^\text{Laughlin}_j$ no longer enters as a common multiplicative factor but instead acts on each orbital separately.
The MHPO-M pair factors $\mathcal{J}_{dij}$ offer a different route.
Rather than relying on an LLL projection, they dress each orbital with a learnable factor that depends on both the orbital and the electron configuration.
Consequently, MHPO-M is not confined to the LLL, which allows it to capture LL mixing.

In recent NNFQH studies, similar learnable orbital factors have also been introduced for a different purpose, namely to enforce exact boundary conditions on the torus~\cite{abouelkomsan_first-principles_2026, graham_non-abelian_2026}.
That approach factorizes each orbital into a product of modified Weierstrass sigma functions with learnable many-body zeros to satisfy magnetic translation constraints, while intentionally avoiding physical priors such as flux attachment or Laughlin correlations.
In MHPO-M, by contrast, the learnable pair factors are introduced specifically to supply additional physical priors without sacrificing flexibility.

Beyond the physical priors discussed above, many others remain to be explored, such as exact rotational symmetry, which is routinely imposed in machine learning force fields~\cite{batatia_mace_2022, tan_high-performance_2026} and in spin-adapted neural network wavefunctions for molecules~\cite{li_saam_2025}.
Achieving this in the present setting is considerably more difficult, as it involves a combinatorial number of choices for the irreducible representations of the generalized orbitals $\varphi_i(\mathbf{r}_j;\{\mathbf{r}_{\ne j}\})$, each of which must be able to carry at least angular momentum $|Q|$ for the electron $j$ to capture LLL physics.
Physical priors can also be imposed through the optimization objective rather than the ansatz.
We analyze angular momentum penalty as one such objective-level prior in the Supplemental Material.

More generally, we demonstrate that physical priors should guide learning without dictating the correlations that the network can learn, and that a good ansatz design should provide the network with the flexibility to represent and optimize them.
As a practical illustration, the MHPO-M* design enables accurate and scalable calculations of FQH systems, opening opportunities to study defects, Wigner crystallization, and the topological order of states at fillings whose ground states remain under debate, such as $\nu=5/2$~\cite{ma_fractional_2024, pakrouski_phase_2015, rezayi_landau_2017}.
The same principle extends naturally beyond spherical geometry, for instance to the torus~\cite{fadon_torus-net_2026, abouelkomsan_first-principles_2026} and to fractional Chern insulators in moiré materials~\cite{wang_fractional_2024, hu_hyperdeterminants_2024, ledwith_vortexability_2023}, where Laughlin-type pair factors remain natural priors on Chern-band orbitals.
We therefore expect that balancing physical priors against learnability will continue to inform the design of more expressive and scalable wavefunction ansatzes.

\begin{acknowledgments}
This work was supported by the National Natural Science Foundation of China under Grant No. 12334003 and No. 125B2083.
We also thank members of the ByteDance Seed Group for their inspiring ideas and encouragement.
\end{acknowledgments}

\bibliography{refs}

\end{document}


\title{Supplemental Material: Neural Network Fractional Quantum Hall States with Learnable Orbital Pair Matrix Factors: Balancing Physical Priors and Flexibility}

\author{Yubing Qian}
\affiliation{School of Physics, Peking University}%
\affiliation{ByteDance Seed}

\author{Weiluo Ren}
\email{renweiluo@bytedance.com}
\affiliation{ByteDance Seed}

\author{Ji Chen}
\email{ji.chen@pku.edu.cn}
\affiliation{School of Physics, Peking University}
\affiliation{Interdisciplinary Institute of Light-Element Quantum Materials and Research Center for Light-Element Advanced Materials, Peking University}
\affiliation{State Key Laboratory of Artificial Microstructure and Mesoscopic Physics and Frontiers Science Center for Nano-Optoelectronics, Peking University}

\maketitle

\begin{table}[tbp]
\caption{\label{tab:params}Neural network and training hyperparameters.}
\begin{tabular}{@{\quad} l @{\qquad} l @{\qquad} l @{\quad}}
\toprule
Category & Parameter & Value \\
\hline
\multirow{5}{*}{Network}
& Number of determinants & 8 \\
& Number of network layers & 4 \\
& Number of attention heads & 4 \\
& Attention dimension & 64 \\
& Fully connected layer dimension & 256 \\
\hline
\multirow{6}{*}{Optimization}
& Learning rate at iteration $t$ & $\operatorname{lr}(t)=r\min\!\left(1,\frac{t}{t_{\mathrm{w}}}\right)\left(1+\frac{t}{t_0}\right)^{-1}$ \\
& Initial learning rate $r$ & $0.05$ \\
& Learning rate warmup $t_{\mathrm{w}}$ & $1000$ \\
& Learning rate delay $t_0$ & $2000$ \\
& Default gradient clipping window & 5.0 (20.0 for MHPO-L and $\nu=1/5$ MHPO) \\
& Default number of iterations & $2\times10^5$ ($4\times10^5$ for $N>9$ systems) \\
\hline
\multirow{3}{*}{Sampling}
& Number of walkers & $4096$ \\
& Decorrelation steps between iterations & $50$ \\
& Number of evaluation steps & $10^4$ \\
\bottomrule
\end{tabular}
\end{table}

\begin{figure}[htb]
    \centering
    \includegraphics{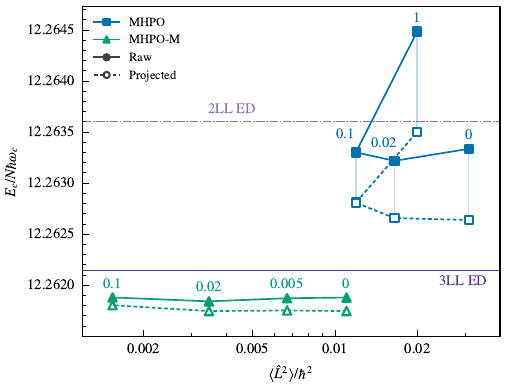}
    \caption{%
    Converged energies (filled markers) and energies projected onto the $L=0$ sector (hollow markers) for MHPO and MHPO-M wavefunctions in the $N=9$, $2Q=24$ system, plotted against the angular momentum penalty strength $\beta$ used during training, with the value of $\beta$ labeled beside each marker.
    The MHPO runs at $\beta=0.1$ and $\beta=1$ are warm-started from the 10k-step $\beta=0$ checkpoint, since training from scratch fails to converge at these penalty strengths.
    The two-LL and three-LL ED energies are indicated by violet dashed lines.
}
    \label{fig:penalty-scaling}
\end{figure}

\section{Angular momentum penalty}

Alternatively, physical guidance can be imposed through the optimization objective.
For instance, adding an angular momentum penalty $\beta\langle \hat L^2\rangle/\hbar^2$, where $\hat L$ is the total-angular-momentum operator and $\beta$ is the penalty strength reported in units of $\hbar\omega_c$, supplies such guidance because the FQH ground states satisfy $\langle \hat L^2 \rangle=0$.
However, since the network has finite capacity, it may struggle to simultaneously achieve low energy and low angular momentum.
Figure~\ref{fig:penalty-scaling} shows the energies obtained with different penalty strengths $\beta$ for the MHPO and MHPO-M wavefunctions in the $\nu=1/3$, $N=9$ system.
To distinguish a lower unprojected energy from an improved state within the target symmetry sector, we also report energies projected onto $L=0$,
\begin{equation}
    E_{L=0}=\frac{\langle \psi_T | \hat H \hat{\mathcal{P}}_{L=0} | \psi_T \rangle}{\langle \psi_T | \hat{\mathcal{P}}_{L=0} | \psi_T \rangle},
\end{equation}
where $\hat{\mathcal{P}}_{L=0}$ is the projection operator onto the $L=0$ sector.
The angular momentum decreases as $\beta$ increases over the range considered, while the unprojected energy is minimized near $\beta=0.02$ for MHPO-M and MHPO.
MHPO-M already has low energy and angular momentum without the penalty, whereas the unprojected MHPO energy benefits more noticeably from moderate guidance.
For both ansatzes, however, $E_{L=0}$ gradually worsens as $\beta$ increases.
Thus, stronger enforcement of this objective-level prior lowers the total angular momentum but does not improve the represented $L=0$ component, even when the unprojected energy improves.

\section{Projection onto the $L=0$ sector}

The projector $\hat{\mathcal{P}}_{L=0}$ is the uniform average over $SU(2)$ rotations,
\begin{equation}
    \hat{\mathcal{P}}_{L=0}=\int_{SU(2)} d\mathcal{R}\, U(\mathcal{R}),
\end{equation}
where $d\mathcal{R}$ is the normalized Haar measure and $U(\mathcal{R})$ rotates every electron coordinate by the same group element: $[U(\mathcal{R})\psi_T](\mathbf{r}_1,\dots,\mathbf{r}_N)=\psi_T(\mathcal{R}^{-1}\mathbf{r}_1,\dots,\mathcal{R}^{-1}\mathbf{r}_N)$.
The projected energy is $E_{L=0}=h_0/p_0$, where the $L=0$ weight $p_0$ and energy numerator $h_0$ are
\begin{equation}
    p_0=\left\langle \mathsf{R} \right\rangle_{\!\{\mathbf{r}\},\mathcal{R}},
    \qquad
    h_0=\left\langle E_{\mathrm{loc}}^{*} \mathsf{R} \right\rangle_{\!\{\mathbf{r}\},\mathcal{R}},
    \qquad
    \mathsf{R}=\frac{\psi_T(\mathcal{R}^{-1}\mathbf{r}_1,\dots,\mathcal{R}^{-1}\mathbf{r}_N)}{\psi_T(\mathbf{r}_1,\dots,\mathbf{r}_N)},
\end{equation}
where $E_{\mathrm{loc}}=(\hat H\psi_T)/\psi_T$ is the local energy and the configurations are sampled from $|\psi_T|^2$.
In practice, these averages are estimated using independent Haar-random $SU(2)$ rotations for each sampled configuration.